\documentclass[amsmath,amssymb,altaffilletter,10pt,aps,pra]{revtex4-1}
\usepackage{graphicx}

\begin{document}

\title{Sub-cycle measurement of intensity correlations in the Terahertz frequency range}%

\author{Ileana-Cristina  Benea-Chelmus}%
\email{ileanab@ethz.ch}
\author{Curdin Maissen}
\author{Giacomo Scalari}
\author{Mattias Beck}
\author{J\'er\^ome Faist}
\email{jerome.faist@phys.ethz.ch}
\affiliation{ETH Zurich, Institute of Quantum Electronics, Auguste-Piccard-Hof 1, Zurich 8093, Switzerland}

\begin{abstract}
The Terahertz frequency range bears intriguing opportunities, beyond very advanced applications in spectroscopy  and matter control.  Peculiar quantum phenomena  are predicted to lead to light emission by non-trivial mechanisms. Typically, such emission mechanisms are unraveled by temporal correlation measurements of photon arrival times, as demonstrated in their pioneering work by Hanbury Brown and Twiss. So far, the Terahertz range misses an experimental implementation of such technique with very good temporal properties and high sensitivity.  In this paper, we propose a room-temperature  scheme to measure photon correlations at THz frequencies based on electro-optic sampling. The temporal resolution of 146~fs is faster than one cycle of oscillation  and the sensitivity is so far limited to $\sim$1500 photons. With this technique, we measure the photon statistics of a THz quantum cascade laser. The proposed measurement scheme allows, in principle, the  measurement of ultrahigh bandwidth photons and paves the way towards THz quantum optics.
\end{abstract}

\maketitle 

\subsection*{Introduction}

The notion of single event detection, and therefore of statistics, is absolutely central in quantum mechanics~\cite{Loudon:2000up}. In fact, the development of quantum optics, first as a tool for fundamental science, now a topic of high technological relevance for quantum cryptography and information processing, is tightly linked to the one of detectors capable of detecting  photons with high efficiency and time resolution~\cite{Hadfield2009}. Measurements of the photon statistics, long restricted to the visible frequency range, are nowadays routinely performed in the telecom frequency range~\cite{Marandi2014} and are being developed in the microwave in the context of circuit quantum electrodynamics~\cite{Corcoles2015,daSilva:2010io}.


Placed at mid-way between the optics and the microwave, the Terahertz frequency region~(100~GHz-10~THz) is potentially very interesting beyond its advanced applications in spectroscopy~\cite{Tonouchi:2007p1411, EiseleM.2014} or matter control~\cite{Kampfrath2013} for the study of linear or nonlinear quantum phenomena at the ultra-strong coupling limit of light-matter interaction~\cite{Ciuti:2005p1558}.  This limit is specifically accessible in this frequency range because of the successful implementation of metallic cavities with strongly subwavelength effective volumes~\cite{Walther:2010p1512} which can be combined with quantum heterostructures in semconductors to form hybrid systems~\cite{Scalari:2012ia,Scalari:2014eh}. In this limit, diverse quantum phenomena have been predicted, such as two-mode squeezing as a consequence of the counter-rotating coupling terms~\cite{Liberato:2007jo}, photon blockade~\cite{Ridolfo:2012dt}, conversion from virtual to real photons~\cite{Ridolfo:2013bi}, and decoupling of light and matter for the deep strong coupling regime~\cite{DeLiberato:2014dy}.

Paramount for the study of these new phenomena is the development of THz detection schemes able to measure efficiently the photon statistics of the quantum source. One possibility to circumvent the absence of materials with bandgaps in the Terahertz is the use of detectors based on intersubband transitions such as Quantum Cascade Detectors or Quantum Well Infrared Photoconductors~\cite{Luo:2005p723} and on single charge sensing in confined heterostructures~\cite{Komiyama:2011cb,Kajihara:2013kj}.  However, these detectors have limited use for quantum statistics measurements due to their high dark count rate (QWIP) or a relatively long dwell time.

In this article, we show that electro-optic sampling~(EOS)~\cite{Wu:1995ec, Gallot:1999ty, Zhao:2002cg} can be employed to measure photon statistics in the Terahertz with unsurpassed sub-cycle resolution~\cite{Gaal:2007iz}, and pave the way for quantum optics experiments in this frequency range.  In electro-optic sampling, the electric field of a THz wave is measured by its nonlinear interaction with a femtosecond probing beam inside a $\chi^{(2)}$ material~(Pockels effect). The short time-resolution characteristic to this technique  has been exploited recently for the direct measurement of zero-point fluctuations of the vacuum electric field at multi-THz frequencies~\cite{Riek23102015}. The THz wave induces a local anisotropic change in the refractive index of the material. As a consequence, the polarization of the linearly polarized probing beam is altered proportionally to the electric field magnitude~\cite{Gallot:1999ty}. The induced refractive index change is typically as small as $\Delta n \approx 10^{-8}$, and lock-in detection is performed, which requires the THz emission to be synchronous with the sampling beam.  Our approach overcomes this restriction of phase-locked THz emission. We validate our technique with the measurement of correlations of a THz Quantum Cascade Laser. Potentially, very high bandwidth photons consisting of only few or even a fraction of a cycle are measurable with this sub-cycle resolving technique.

\subsection*{Experimental setup}

Often, the quantum nature of a source and its emission dynamics are reflected in the temporal distribution of the emitted photons. Typically, a single photon source, such as a single - artificial -  atom, emits one single photon at a time~(anti-bunching), while, on the contrary, a thermal source emits many photons in bunches~(bunching). Experimentally, this property is retrieved from a measurement of field and intensity correlation functions~\cite{Boitier:2009p1514, daSilva:2010io}, which have the following mathemalical expressions~\cite{Loudon:2000up}.
$$g^{1} (\tau)= \frac{ \langle \mathcal{E}^{-}_{THz}(t) \mathcal{E}^{+}_{THz}(t+\tau) \rangle_{t} }{\sqrt{\langle \mathcal{E}^{-}_{THz}(t)  \mathcal{E}^{+}_{THz}(t)  \rangle_{t} \langle \mathcal{E}^{-}_{THz}(t+\tau)  \mathcal{E}^{+}_{THz}(t+\tau) \rangle_{t}}}$$
 $$g^{2}(\tau) = \frac{ \langle \mathcal{E}^{-}_{THz}(t)\mathcal{E}^{-}_{THz}(t+\tau) \mathcal{E}^{+}_{THz}(t)\mathcal{E}^{+}_{THz}(t+\tau) \rangle_{t} }{\langle  \mathcal{E}^{-}_{THz}(t)  \mathcal{E}^{+}_{THz}(t) \rangle_{t} \langle \mathcal{E}^{-}_{THz}(t+\tau)  \mathcal{E}^{+}_{THz}(t+\tau) \rangle_{t}}.$$

Here,  $g^{1}(\tau)$  represents the degree of first order coherence and gives the spectral composition of the THz emission, as in a Fourier Transform Infrared Spectrometer~(FTIR). $g^{2}(\tau)$ is the degree of second order coherence.

\begin{figure}
  \centering
  \includegraphics[width=\textwidth]{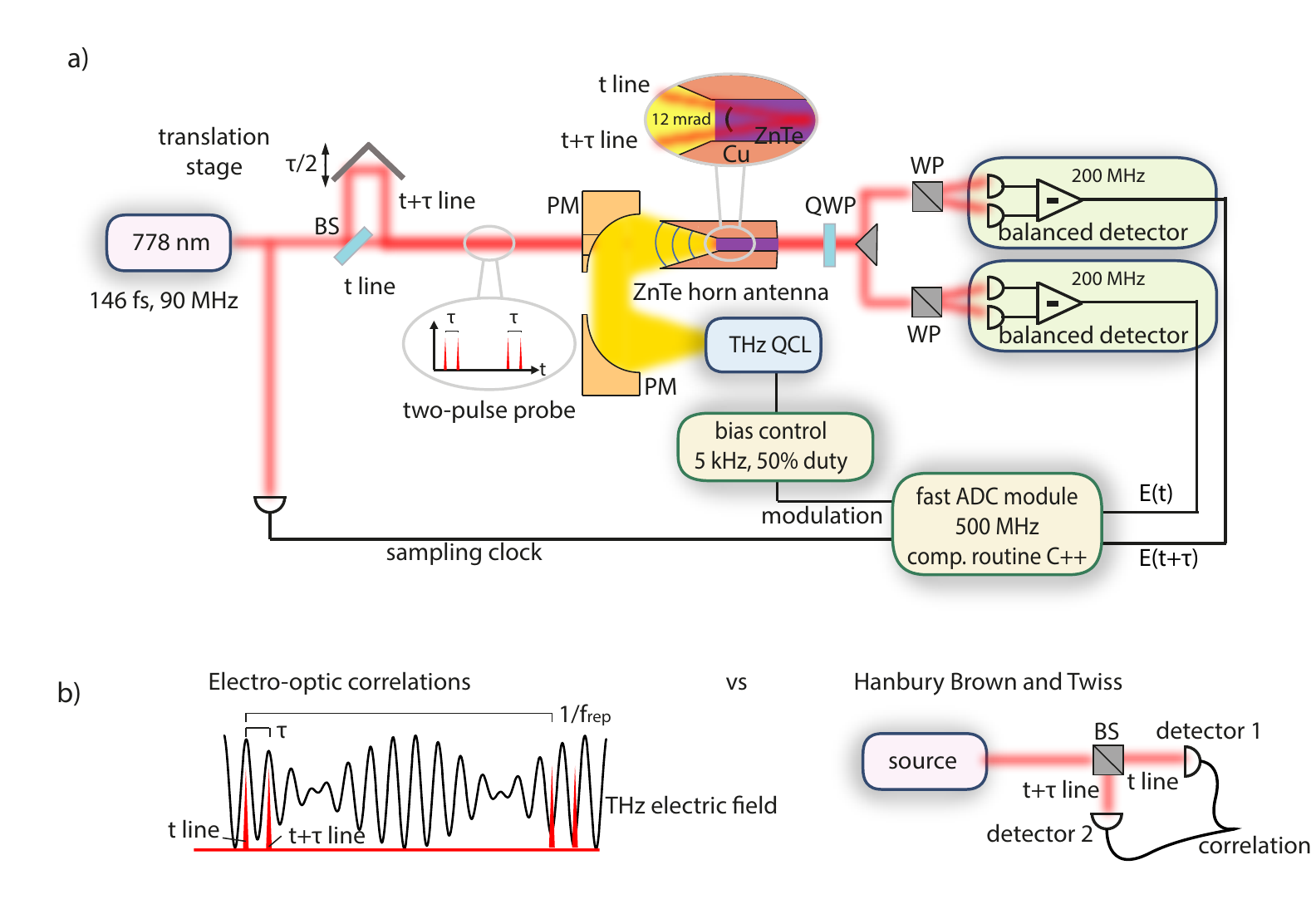}
  \caption{Sketch of the setup to measure field and intensity correlations in the THz range by electro-optic sampling. \textbf{a} We exploit the electro-optic effect combined with a two-pulse probe scheme and ultra-fast acquisition at the repetition rate of the sampling laser ($f_{rep}$~=~90 MHz). The electric field of the THz wave under investigation, here originating from a free-running THz Quantum Cascade Laser, is sampled at two mutually delayed time-points~(t and t+$\tau$), adjusted with a delay stage. The measured electric field values are recorded at the rate of arrival of laser pulses with an analog-to-digital converter~(ADC) and further processed with a C++ computation routine to measure $g^{(1)}(\tau)$ and $g^{(2)}(\tau)$. For increased SNR, the field and intensity correlation terms are measured via a lock-in technique~(5~kHz, 50~\% duty cycle). \textbf{b} Comparison between electro-optic correlation scheme and Hanbury Brown and Twiss approach. Two-pulse electro-optic sampling allows for the measurement of correlations without any beam splitters and with sub-cycle resolution, which is limited by the width of the fs probing pulse to 146~fs.}
  \label{fig:setup} 
\end{figure}

We exploit electro-optic sampling to measure the degree of first and second order coherence of free-running THz beams at sub-cycle timescales. The temporal resolution is limited by the width of the femtosecond laser pulse to 146~fs. A sketch of the setup is shown in figure~\ref{fig:setup}. Two probing beams with a mutually adjustable delay $\tau$ sample the electric field of the THz wave and allow for the investigation of time-related correlations~($\tau=\pm140~ps$, refer to figure~\ref{fig:setup}b). They originate from the same main oscillator and are separated with a beamsplitter~(BS). They enter at a very shallow angle of $\sim12$~mrad inside the nonlinear crystal where they interact with the THz wave and are finally spatially separated at a distance d~=~1~m after the crystal. A key building block of our setup, and new with respect to already reported EOS setups, is the horn antenna which hosts the ZnTe crystal. It is engineered to exhibit an enhanced electro-optic effect due to favorable coherence properties between the near-IR probing beam at 778~nm and the THz~source at 2.3~THz with an theoretical coherence length of 8~mm. In addition, the horn antenna coupled copper waveguide collects the  wave efficiently~(matched for broadband collection), enhances the electric field in vertical direction by a factor of $\sim$1.5 by soft waveguiding and  prevents the divergence of the THz beam within the interaction crystal~( f-number N=1 of parabolic mirror PM, EFL = 3"). For details on the design and frequency properties, consult the methods section. A major advantage of our system is that it operates at room temperature. Because of the residual electro-optical crystal absorption and thermal photon background, however, it also prevents single photon counting. Enclosing the experiment in a cryostat would overcome these limitations.  

\begin{figure}
  \centering
  \includegraphics[width=\textwidth]{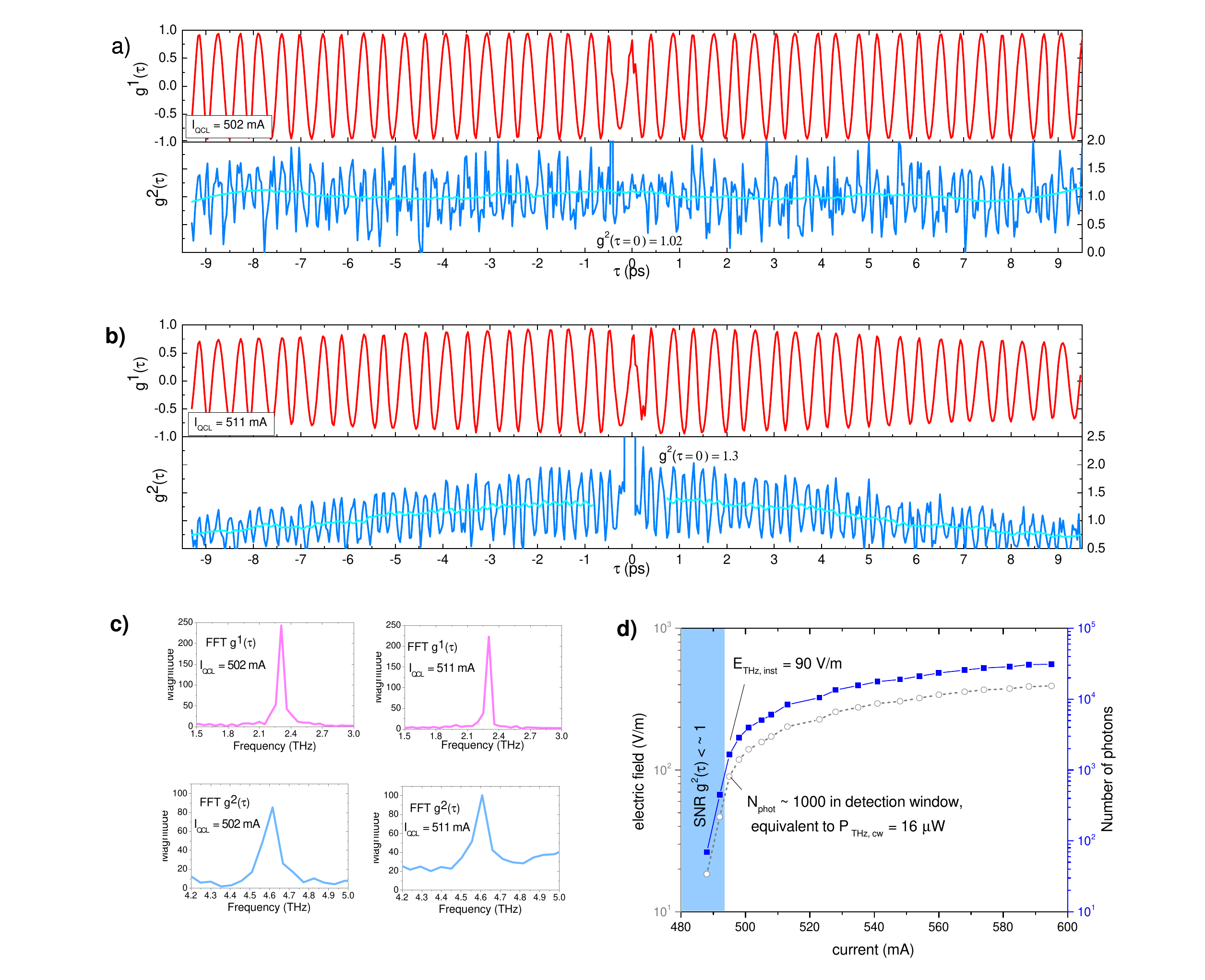}
  \caption{Sub-cycle correlation results of a THz Quantum Cascade Laser lasing around 2.3 THz and setup sensitivity. a) $g^{1}(\tau)$ and $g^{2}(\tau)$ in the singlemode emission regime. The few-cycle average of  $g^{2}(\tau=0)=1.02$  corresponds to the expected theoretical value of a pure coherent state~($g^{2}(\tau=0)=1$). b) $g^{1}(\tau)$ and $g^{2}(\tau)$ in the multimode emission regime. The few-cycle average of  $g^{2}(\tau=0)=1.3$ is due to multimode emission. c) Fourier transformations of aforementioned $g^{1}(\tau)$ and $g^{2}(\tau)$ reveal laser operation around 2.3~THz and characteristic components of $g^{2}(\tau)$ at double frequency 4.6~THz because of sub-cycle temporal resolution. d) Emission characteristics of the THz QCL under investigation, given in electric field or number of photons in the detection time-window indicate a threshold around 495~mA. Below the limit number of photons of $\sim$1500, the signal-to-noise ratio of the $g^{2}(\tau)$ measurement becomes inferior to one.}
  \label{fig:correlations_expanded}
\end{figure}

Our implementation of correlation measurements is based on the sampling of the electric field and intensity at mutually delayed time-points and does not require any beamsplitters for the THz, contrary to the well-established Hanbury Brown and Twiss setup~\cite{BROWN1956}, as shown in figure~\ref{fig:setup}b). It can be easily extended to measure even higher order correlations ($g^{n}(\tau), n>2$) by adding additional probing beams, without any intensity loss at the detectors. Hereby, processes with non-Gaussian statistics could be investigated~\cite{Stevens:2010kra,Chopra:1973js}. Additionally, even long-time delay coherence properties on the order of nanoseconds to microseconds ($10^{3}-10^{6}$ times longer than the period of one single oscillation of the electric field) can be easily  measured, since an additional delay of $\tau = N \times~11~ns$ can be added by computing the correlation between asynchronous laser pulses.

The acquisition of the measured electric field is performed in real-time on the time-scale of every individual sampling pulse~(90~MHz) by two balanced detectors~(bandwidth of 200 MHz, faster than the repetition rate of the laser).  This secures the sub-cycle temporal resolution for the correlation measurements. The double-beam configuration acts as self-referencing  and translates the time averaging of terms such as $ \langle \mathcal{E}_{THz}(t) \mathcal{E}_{THz}(t+\tau) \rangle_{t} $, $ \langle \mathcal{E}_{THz}(t)^{2} \mathcal{E}_{THz}(t+\tau)^{2} \rangle_{t} $ into an averaging over the whole set of acquired data~($ \langle \rangle_{t} = \langle \rangle_{i}, \forall i$~measured), since the phase of the electric field is random.  A simple and fast C++ routine performs the direct and online computation of  $g^{1}(\tau)$ and $g^{2}(\tau)$. The detection is shot-noise limited with a noise equivalent field per single pulse of 600~V/m at the emission frequency of 2.3~THz, with an effective coherence length of 0.5~mm~(as developed in the Methods section). For an increased signal-to-noise ratio, the THz emission is modulated at 5~kHz, 50\%~duty cycle and a digital lock-in type of acquisition is implemented for all time-averaged terms, by using the oscillator repetition rate as sampling clock at 90~MHz. For a full description of the acquisition algorithm, see Methods.

\subsection*{Sub-cycle photon correlations}

We demonstrate the capabilities of our setup with the measurement of the coherence properties of a THz Quantum Cascade Laser~(QCL) over its full dynamical range, including the threshold. The fundamental linewidth of such a  system was measured recently~\cite{Ravaro:2012bn,Vitiello:2012ds}, demonstrating very narrow (100's Hz) linewidth that are Schawlow-Townes limited, as expected~\cite{Faist:2013td}. Nevertheless, the possible role of an additional broadening caused by amplified thermal emission remains experimentally and theoretically open~\cite{Yamanishi:2012iv}. A measurement of $g^{(2)}(\tau)$ reveals information about the current dependent photon statistics of the source.

The device was designed for operation at low frequency (2.3 THz) using a four quantum well design~\cite{Turcinkova:2011jp}. It is implemented with a Fabry-Perot cavity and fitted with a Silicon lens to enhance the light extraction. The laser threshold is identified close to 495~mA at T=25~K for ridge dimensions of 150~$\mu$m~x~1000~$\mu$m~x~16.6~$\mu$m. 

We performed measurements of field and intensity correlations, $g^{(1)}(\tau)$ and  $g^{(2)}(\tau)$, at distinct operation currents above threshold. In figure~\ref{fig:correlations_expanded} we report  sub-cycle results of  $g^{(1)}(\tau)$ and  $g^{(2)}(\tau)$ in the singlemode emission regime, at 502~mA (a)), and in the multimode emission regime, at 511 mA (b)). We identify several expected results. Firstly, the frequency spectrum of the field autocorrelation reveals the operation frequency as expected around 2.3~THz. Secondly, the sub-cycle nature of the measurement of $g^{(2)}(\tau)$ is reflected in the oscillatory character with Fourier components at double the emission frequencies. Thirdly, the few-cycle average value of $g^{(2)}(\tau=0)$ of 1.02 at 502~mA is consistent with pure coherent state emission whereas of 1.3 at 511~mA is consistent with multimode emission. Our detection time-resolution of  $\delta t_{fs}$=146~fs corresponds to a Nyquist limit of 3.37~THz, and is clearly satifactory for the investigated device emitting at 2.3~THz.

The dependence of the electric field on the input current clearly indicates the presence of lasing threshold around 495~mA~(figure~\ref{fig:correlations_expanded}d), at which point the electric field takes values of 50-90~V/m, roughly 10 times smaller than the noise equivalent field per pulse~(600~V/m). The effective number of photons concentrated in the ultra-short detection time window of  $\Delta t_{fs}$=146~fs amounts to roughly 1500. The corresponding average continuous wave output power scales therefore with the duty cycle represented by $\frac{1}{\Delta t_{fs}}$  according to $P_{THz, cw} = \frac{N h\nu}{\Delta t_{fs}}$ amounts to approximately $16~\mu$W.

\subsection*{Photon statistics for the full dynamic range}

\begin{figure}
  \centering
  \includegraphics[width=\textwidth]{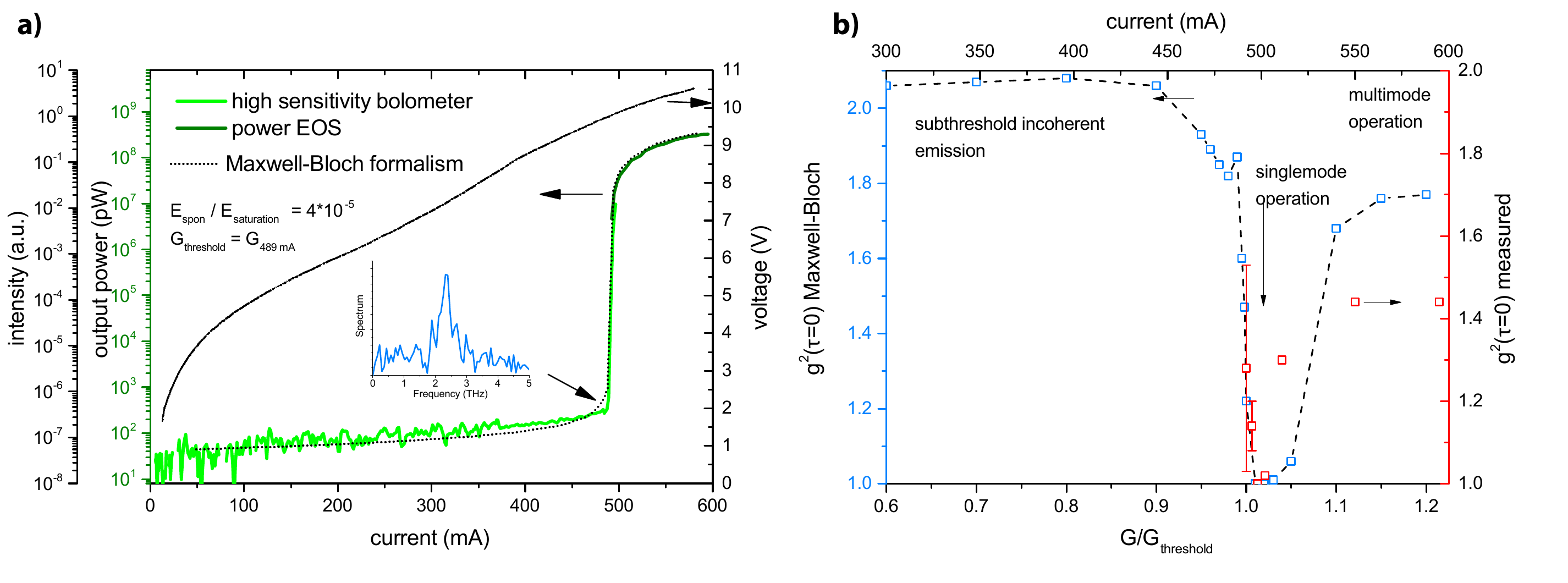}
  \caption{Comparison experimental results and theoretical prediction by Maxwell-Bloch equations. a) Excellent agreement of the laser dynamics over full current range. b)Theoretical~(blue) and experimental~(red) intensity correlation results for the full dynamical range of the THz QCL. $g^{(2)}(\tau = 0)$  as a function of input current shows good agreement with theory: coherent emission at threshold, multimode emission at high input current and onset of incoherent fluctuations at lasing threshold.}
  \label{fig:g2threshold}
\end{figure}

A laser changes its photon statistics at threshold, switching from spontaneous emission below lasing threshold with a  $g^{(2)}(\tau = 0) = 2$ towards coherent radiation above threshold with  $g^{(2)}(\tau = 0) = 1$, for a single mode device. At threshold, a transition between these two values is expected. We performed such experimental study on our THz QCL and found multimode emission at high operating current with a $g^{(2)}(\tau = 0) = 1.44$, coherent emission into one single lasing mode only in a short current range above threshold, with a $g^{(2)}(\tau = 0) = 1$, and an increased  $g^{(2)}(\tau = 0) = 1.28$ in the threshold region~(refer to figure~\ref{fig:g2threshold}). 

We complement our experimental study with theoretical predictions on the photon statistics and the intra-cavity photon field in the whole range of operation of the laser. The performed theoretical study is based on a Maxwell-Bloch formalism combined with modal decomposition and has been extensively discussed in Ref~\cite{khurgin:2013um,Villares:2015ho}. We add to the existing model the spontaneous emission, which is the limiting quantum noise. The latter is modeled as a Markoffian random process with an amplitude corresponding to one single photon inside the laser cavity. The solution of this dynamic model describes the temporal evolution of the amplitudes of all the modes inside a laser cavity at any given pumping parameter, described by the gain parameter $G$. This parameter plays the same role as the driving current in the experimental realization.

The Maxwell-Bloch approach has been used to predict the output power - current characteristics,  as well as the second order degree of coherence  over the full dynamic range of the laser. For our computation, we used the following parameters:  $\tau_{coh} = 0.5~ps$ (equivalent to 2.6~meV bandwidth), $\tau_{up}=5~ps$, $\tau_{photon}=35~ps$ (corresponds to mirror reflectivity of 70\%), $\tau_{round-trip} = 4~ps$ (corresponds to a laser cavity of 1~mm), a GVD of $6.24\cdot10^{5} fs^{2}/mm$~\cite{Rosch:2014ft}, a dipole of $z_{12}=7~nm$ and a ratio of single photon field to saturation field of $4\cdot10^{-5}$. The latter represents the electric field amplitude of a single photon inside the laser cavity decribed by $A_{single photon}=\sqrt{2\pi\hbar\nu/2\epsilon_{0}\epsilon_{r}V_{cavity}}$ normalized to the saturation amplitude as required by the Maxwell-Bloch formalism. Experimentally, its square is an rough indicator of the ratio of the intracavity photon field below threshold to the photon field above threshold. The current-output power characteristics is reported in figure~\ref{fig:g2threshold}a, together with  high-sensitivity bolometric power measurements and electro-optic sampling measurements. We find excellent agreement between predicted and measured data. The threshold is characterized by a power drop of more than five decades whithin a very limited range of input current~(roughly 10~mA). We highlight here that no parameter fitting has been performed.

In addition, we report on the predicted values of $g^{2}(\tau=0)$ as a function of input current in figure~\ref{fig:g2threshold}. We find an agreement between the model and the measurement for several aspects: the short current range of single mode emission with a  $g^{2}(\tau=0)=1$ around the lasing threshold and the onset of multimode emission at currents above 505~mA. It overestimates $g^{2}(\tau=0)$ in the regime of multimode emission. We emphasize here that, the latter is highly susceptible to the exact amplitudes of the modes, which can be easily influenced by any external feedback, imperfections in the processing or any other reminiscent cavity effect in our setup. In addition, the measurement indicates that the photon statistics change at threshold~(increase up to 1.28). However, a discrepancy still persists for the case where at threshold, the value of $g^{2}(\tau=0)$ considerably increases above 1,  between the effective number of photons measured and the one expected. Even though this effect is not fully understood yet, it might point towards drifts in the output power of the laser due to heating rather than quantum noise due to spontaneous emission. Amplified thermal photons might as well play a role.

\subsection*{Conclusion and outlook}
%

In conclusion, we have demonstrated the successful implementation of a THz detection scheme which gives access to first and second order coherence with a time resolution of 146~fs and a sensitivity limited to $\sim$1500 photons. We apply this technique to a THz Quantum Cascade Laser and measure its statistics over the full dynamical range, and the threshold. Such a technique might be applied to the newly developed THz QCL combs to help clarify the time structure of their output intensity~\cite{Burghoff:2014hpa,Rosch:2014ft}.

Moreover,  reaching the ultimate goal of single photon sensitivity, is of high technological importance to the field of THz quantum optics systems. We believe that this limit can be reached with few technical improvements in the detection scheme and the probing beam power. Recent success in making single photons interact in nonlinear optical media strengthen our ambition and have already changed the general perception that nonlinear effects can only be observed at high photon intensities and are vanishingly small for single photons~\cite{Chang2014}. As demonstrated, our averaged sensitivity limit corresponds to an equivalent electric field of 90~V/m detected along an effective length of 0.5~mm. The single photon detectivity demands therefore for a signal  enhancement of 40. This can be gained by several improvements:  polarization control in a Brewster angle scheme as proposed in reference~\cite{Ahmed:detenh}; cryogenical cooling of the ZnTe crystal narrows absorption lines and is expected to increase the coherence length to several milimiters. Further improvements can be gained by confinement of the THz field inside the nonlinear waveguide at dimensions on the order of the probed wavelength inside the material or by the interplay of probe wavelength and detected frequency which create coherence windows in any frequency range from 100~GHz to 3~THz.

\subsection*{Acknowledgement}
The presented work is supported in part the ERC project MUSIC as well as by the NCCR  Quantum Science and Technology.

\bibliography{bibtex-library}
\end{document}